# The Rapid Transient Surveyor

C. Baranec[a*], J. R. Lu[a†], S. A. Wright[b], J. Tonry[a], R. B. Tully[a], I. Szapudi[a], M. Takamiya[c], L. Hunter[d], R. Riddle[e], S. Chen[f], & M. Chun[a]

[a]Institute for Astronomy, University of Hawai'i at Mānoa, Hilo, HI 96720-2700, USA; [b]Center for Astrophysics and Space Sciences, University of California, San Diego, La Jolla, CA 90037, USA; [c]Department of Physics and Astronomy, University of Hawai'i at Hilo, Hilo, HI 96720, USA; [d]Institute for Science & Engineer Educators, University of California, Santa Cruz, CA 95064, USA; [e]Division of Physics, Mathematics and Astronomy, California Institute of Technology, Pasadena, CA 91125, USA, [f]Dunlap Institute for Astronomy & Astrophysics, University of Toronto, M5S 3H4, Canada, [†]Now at: Astronomy Department, University of California, Berkeley, CA 94720, USA

## ABSTRACT

The Rapid Transient Surveyor (RTS) is a proposed rapid-response, high-cadence adaptive optics (AO) facility for the UH 2.2-m telescope on Maunakea. RTS will uniquely address the need for high-acuity and sensitive near-infrared spectral follow-up observations of tens of thousands of objects in mere months by combining an excellent observing site, unmatched robotic observational efficiency, and an AO system that significantly increases both sensitivity and spatial resolving power. We will initially use RTS to obtain the infrared spectra of ~4,000 Type Ia supernovae identified by the Asteroid Terrestrial-Impact Last Alert System over a two year period that will be crucial to precisely measuring distances and mapping the distribution of dark matter in the $z < 0.1$ universe. RTS will comprise an upgraded version of the Robo-AO laser AO system and will respond quickly to target-of-opportunity events, minimizing the time between discovery and characterization. RTS will acquire simultaneous-multicolor images with an acuity of 0.07-0.10" across the entire visible spectrum (20% i′-band Strehl in median conditions) and <0.16" in the near infrared, and will detect companions at 0.5" at contrast ratio of ~500. The system will include a high-efficiency prism integral field unit spectrograph: $R = 70-140$ over a total bandpass of 840-1830nm with an 8.7" by 6.0" field of view (0.15" spaxels). The AO correction boosts the infrared point-source sensitivity of the spectrograph against the sky background by a factor of seven for faint targets, giving the UH 2.2-m the H-band sensitivity of a 5.7-m telescope without AO.

## 1. INTRODUCTION

The next decade of astronomy will be dominated by large area surveys (see [1]). Ground-based optical transient surveys, e.g., LSST and ZTF, and space-based exoplanet, supernova, and lensing surveys such as TESS and WFIRST will join the Gaia all-sky astrometric survey in producing a flood of data that will enable leaps in our understanding of the universe. There is a critical need for further characterization of these discoveries through high angular resolution images, deeper images, spectra, or observations at different cadences or periods than the main surveys [2]. Such follow-up characterization must be well matched to the particular surveys, and requires sufficient additional observing resources and time to cover the extensive number of targets.

The Rapid Transient Surveyor (RTS) will uniquely address this need by combining an excellent observing site, the unmatched efficiency of the prototype Robo-AO system, and copious available time on the UH 2.2-m telescope. This will enable the high-acuity and sensitive spectral follow-up of tens of thousands of objects in mere months. The system will respond quickly to target-of-opportunity events, minimizing the time between discovery and characterization, and will interleave different science programs with its intelligent queue. We are also planning on eventually making RTS available to others for both standalone surveys and for rapid follow-up of other surveys, using the same trigger system developed for ATLAS, and partnering with projects such as the ANTARES event broker [3] to streamline requests.

Our primary RTS science goal is to map the dark matter distribution in the $z < 0.1$ local universe with ten times better accuracy and precision than previous experiments. We will use the Asteroid Terrestrial-Impact Last Alert System (ATLAS [4]; PI J. Tonry) to discover several thousand SNIa per year using automatic detection routines, and follow discoveries with triggered observations with the RTS spectrograph. ATLAS will measure SNIa peak brightness, and decline rates, and the RTS observations will measure reddening by dust, confirm SN types and confirm redshifts to identify host galaxies. This unique combination of automated detection and characterization of astrophysical transients during a sustained observing campaign will yield the necessary statistics to precisely map dark matter in the local universe.

*baranec@hawaii.edu; 1-808-932-2318; http://high-res.org; http://robo-ao.org

## 2. THE RAPID TRANSIENT SURVEYOR INSTRUMENT

RTS will combines an adaptive optics (AO) system with an infrared integral field spectrograph (IFS) mounted to a robotized 2.2-m telescope at the excellent Maunakea observing site in Hawaii. In addition to the typical benefit of improved angular resolution, AO is crucial in concentrating the light of faint targets against the bright infrared sky, significantly shortening exposure times. The intelligent target queue system will interleave requests for transient observations with other science programs and external requests, to maximize the scientific throughput.

### 2.1 Adaptive optics system

The RTS AO system will be the functionally similar to the prototype Robo-AO system (Figure 1) [5,6] recently transferred from the Palomar 1.5-m to the Kitt Peak 2.1-m telescope [7]. The new RTS AO system will exploit a larger telescope aperture and Maunakea's excellent observing conditions to obtain sharper and higher-contrast images at even shorter wavelengths. As described in §4.1, we take advantage of this exquisite image sharpening to boost the infrared point-source sensitivity of the IFS which significantly increases its observational efficiency. Just as importantly, the new RTS system will be permanently mounted and will be available year round to support the long-duration observing campaigns necessary to execute the science programs described herein.

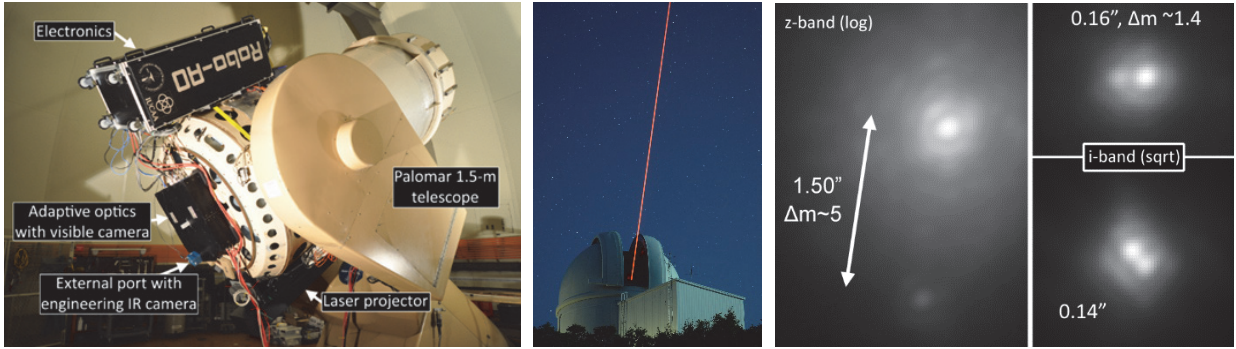

**Figure 1. Left:** The prototype Robo-AO system on the 1.5-m telescope at Palomar. **Center:** The UV laser propagating on sky. Although invisible to the human eye, standard digital cameras can see the UV light with their UV blocking filters removed. **Right:** Example Robo-AO corrected visible light images of binary stars.

### 2.2 Laser guide star

In order to make corrections anywhere in the sky, the RTS AO system is equipped with a laser guide star to measure the atmospheric blurring. We will copy the UV laser and laser projector that is used with the prototype AO system but with two modifications. To compensate for the observatory's higher elevation, a longer range-gate (667 m vs. 375 m) will be used, necessitating a periscope to be added to the end of the telescope tube to jog the beam on-axis to reduce perspective elongation of the laser. While the laser projector pointing is stable to the level correctable by an internal uplink tip-tilt mirror, the periscope will include active pointing to compensate for deterministic mechanical flexure. We will also implement passive (e.g., using a carbon fiber or invar honeycomb breadboard) and/or active controls on the laser projector focus to compensate for the temperature dependent focus drift found with the prototype.

Because the UV laser is invisible to the naked eye, it has been approved for use without human spotters by the Federal Aviation Authority; however, coordination with U.S. Strategic Command is still necessary to avoid illuminating critical space assets. Instead of obtaining clearance for individual targets, the prototype operated under a system whereby we requested clearance for azimuth and elevation ranges (each 2.5°×2.5°) over the entire sky above 40° elevation, ensuring that at any given time there are targets available to observe with the laser. This also allows us to observe new targets, such as newly discovered transient events, with the laser on the fly. This procedure was used at Palomar, is being used at Kitt Peak, and is planned for use by RTS and the Keck LGS AO systems on Maunakea.

## 2.3 Cassegrain mounted adaptive optics system

The adaptive optics system and science cameras will reside within a Cassegrain mounted structure of approximate dimensions 1 m × 1 m × 0.2 m, and will be conceptually similar to the layout presented in Figure 2. Light from the telescope will enter the instrument and be intercepted by a fold mirror which directs a 2 arc minute diameter field to a dual off-axis parabolic (OAP) mirror relay. The first fold mirror will be on a linear motorized stage that can be moved out of the way to reveal a calibration source that matches the 2.2-m telescope focal ratio and exit pupil position while simultaneously mimicking an ultraviolet laser source at 10 km and a combination flat field/point source at infinity. The later will use a beam-combiner to optically feed an unresolved single-mode fiber tip, or the exit aperture of a small integrating sphere with inputs from a thermal and wavelength calibration sources.

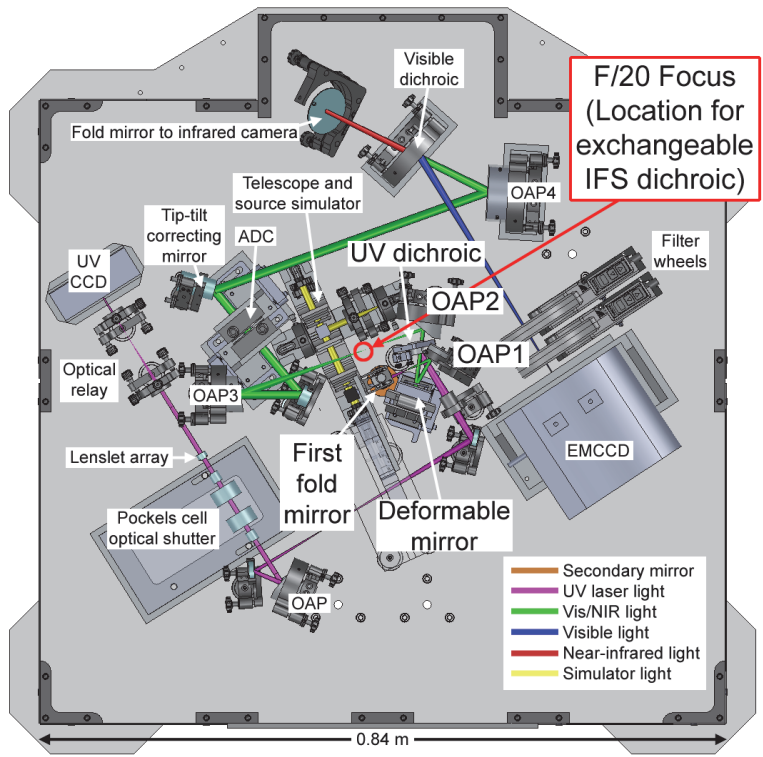

The first OAP will image the telescope pupil onto the deformable mirror. After reflection off the deformable mirror, the UV laser light will be selected off with a dichroic mirror and sent to a UV optimized wavefront sensor. The visible and infrared light will pass through the dichroic and will be refocused by another OAP (#2). The instrument will have two selectable configurations: a mode similar to the prototype where there is a visible and infrared camera sharing the F/41 output of the dual OAP relay, and a second mode where the visible light still goes to the same visible camera, but, the infrared light is instead picked off after the first relay at F/20 and sent to the IFS (Figure 5). To select between the two modes, there will be a mirror selection mechanism before the focus of OAP#2 that will either a) reflect all of the light to the second relay or b) a dichroic that reflects visible light and transmits $\lambda > 830$ nm to the infrared spectrograph. By selecting the infrared light for the spectrograph after the first relay, we will be able to minimize losses of the adaptive optics system, keeping the throughput to >75% (4 reflections and 4 transmissions, based on measurements of prototype component %R/%T). Note that the spectrograph will have its own tip-tilt corrector that will be slaved to the tip-tilt corrector in the second OAP relay.

**Figure 2.** Solid model of the prototype adaptive optics support structure. The new layout will include an exchangeable dichroic at the F/20 focus that will supply the IFS with an AO-corrected high-throughput (>75%) beam. The increased back focal distance of the UH 2.2-m telescope (313mm vs. 131mm at Palomar) will ease the packaging of optics within the RTS support structure.

When not using the IFS, all of the light is relayed by a second OAP relay which includes a tip-tilt corrector and an atmospheric dispersion corrector (for 425nm < $\lambda$ < 1.8μm). The final relay element will create a telecentric F/41 beam which will be split by a fixed dichroic mirror at $\lambda$ = 830 nm. Each channel will have an imaging camera (§2.6) which can double as a tip-tilt sensor. The visible channel will have a fold mirror on a linear stage which can fold the beam out of the instrument to ports where future external instruments can be mounted (e.g., an eyepiece [8], student-built cameras/spectrographs, etc.).

## 2.4 Upgraded UV optimized wavefront sensor

Compared to the prototype, RTS will more finely sample the wavefront with 11.6 cm$^2$ subapertures (vs. 13.6 cm$^2$), that increases the pupil sampling to 19×19 subapertures. We will use a UV optimized FirstLight OCAM2K EMCCD camera (240$^2$ format, with 0.3e$^-$ of read noise and 1.4 excess noise factor at a 2kHz full frame rate – a standard version of this camera is being used at Subaru). The detector is a special E2V CCD220 which will have the same UV anti-reflection coating as our previous E2V CCD39 and will have >50% QE at the laser wavelength. We will bin the pixels by a factor of 2 and use 6×6 binned pixels to calculate the slopes in each subaperture, leading to a much more linear response than a simple bi-cell [9]. We will operate at a frame rate of 1.5kHz while reading out the frame transfer area at the maximum of 3.7kHz to reduce latency. Even with the greater number of pixels being read, and faster frame rate, the EM gain allows us to maintain a higher slope measurement SNR.

Range gating of the pulsed laser, to block Rayleigh scattered light outside the focused beam waist, will still be accomplished with a Pockels cell between two crossed linear polarizers.

### 2.5 Upgraded deformable mirror

To match the new wavefront sensor, we will upgrade the deformable mirror from a Boston Micromachines 12×12 actuator continuous face-sheet mirror to a 492 actuator version (we use 20 of 24 actuators across a circular aperture). The same 3.5 μm stroke is more than adequate in all but the most extreme seeing conditions. The new deformable mirror has a total latency and response of <50 μs, approximately half of the previous mirror. We will similarly adopt a Fried geometry and use a subset of actuators within a circle, 20 actuators wide, for wavefront correction. The previous challenge of packaging optics around a deformable mirror with a 4.4 mm diameter pupil will be alleviated with the new mirror with its larger 7.6 mm diameter pupil.

### 2.6 Imaging cameras / tip-tilt sensors

The adaptive optics system will have both visible and infrared cameras which can be used for imaging and/or tip-tilt sensing (summary in table 1). We will use an Andor iXon Ultra 888 cameras for our visible channel. This camera uses the same E2V CCD201 electron-multiplying CCD as the prototype Robo-AO Andor 888 camera but with two major improvements; new electronics are optimized for faster readout, so the cameras can read out the full frame at 26 Hz as opposed to 8.6 Hz – providing better post-facto image motion correction on brighter targets. There is also a new USB 3.0 interface which obviates the need of custom Andor camera interface cards in the control computer.

| Detector | Format | Field | Pix.Scale | ReadNoise | Full frame rate | Tip-tilt rate | Initial filters |
|---|---|---|---|---|---|---|---|
| EMCCD | $1024^2$ | 31″×31″ | 0.030″ | <1e- | up to 26 Hz | to 500 Hz | g', r', i', z', Hα, r'+i'+z' |
| SAPHIRA | 320×256 | 17″×14″ | 0.055″ | <1e- | up to 400 Hz | to 8 kHz | Y, J, H, FeII, Y+J+H |

**Table 1.** Format and characteristics of imaging and tip-tilt detectors within the RTS AO system.

We will use a Selex ES SAPHIRA detector in a GL Scientific cryostat for the infrared camera. These detectors were originally developed by ESO and Selex for the GRAVITY instrument on the VLTI [10] and have been adopted by ESO for future AO wavefront sensing applications. Don Hall has been funded by the NSF to further develop these 320×256 devices for infrared tip-tilt sensing and by NASA for potential space applications as they are capable of very fast readouts, >400 Hz full frame with 32 parallel outputs, and very low noise, <1e-, via avalanche amplification of the signal in the HgCdTe before being read out. In summer 2014, we tested an early device as a simultaneous tip-tip sensor and science camera [11]. For the RTS AO system, we will incorporate one of the newly delivered science grade SAPHIRA detectors (AR-coated, sensitive from 0.8-1.8 μm, and with only a few bad pixels) into a cryostat funded by the Office of Naval Research. To simplify the camera, we will install a cold, λ > 1.8 μm blocking filter immediately in front of the detector and use a warm filter wheel directly in front of the entrance window. As a future upgrade path, the SAPHIRA cryostat will be able to accommodate much larger format detectors as they become available.

### 2.7 Robotic software

The RTS AO system will reuse the robotic software developed for the prototype (fully detailed in [12]), with continuing improvements to reliability and functionality being made during the deployment to Kitt Peak. A single computer commands the AO system, the laser guide star, the science cameras, the telescope, and other instrument functions. The easy-to-modify C++ software was designed from the outset to be very modular: the software to control each hardware subsystem was developed as a set of individual modules and small standalone test programs have been created to test each of the hardware interfaces. This modular design allows the individual subsystems to be stacked together into larger modules, which are then controlled by the master robotic control system.

The robotic control system executes tasks that have previously been performed manually; e.g., to start an observation, the central robotic control daemon will point the telescope, move the science filter wheel, and configure the science camera, laser, and AO system, during the telescope slew. After settling, an automated laser acquisition process executes, and an observation begins. A redundant safety system manages laser propagation onto the sky, and stops laser operations if any errors occur. After completion of an observation the control system will query the intelligent queue for the next target to observe. An analysis of the logs over the last year at Palomar shows that the instrument takes ~30 seconds to set up the instrument for each observation (excluding slew time).

The intelligent queue [13] is able to pick from targets in a directory structure organized by scientific program, with observation parameters defined inside of XML files. All targets from all programs are initially considered "available" and the queue uses an optimization routine based on priority, slew time and cutoffs (including time, elevation, or hour angle windows, and whether it is safe to use the laser for the duration of the observation) to determine the next target to observe. Adding targets to the queue is as simple as creating a new XML file in the queue directory; adding transient targets will work in the same way, and can be added with time dependent scientific priority to aid in observing SNIa near their peak luminosity. For targets that require immediate observations, new XML files can be added as highest priority and will be executed next in the queue presuming other safety checks including telescope pointing limits and laser clearance is passed.

To have the RTS AO system work as the prototype does now, we will write new modules for the upgraded pieces of hardware: the visible, infrared and wavefront sensor cameras; the deformable mirror; and the UH 2.2-m telescope control system. An existing fast-frame rate visible light camera data reduction pipeline currently produces fully reduced images in the morning after the data are acquired. We will create new automatic data reduction pipeline for the infrared camera, as well as handle long-integration data from each of the cameras. The spectrograph data pipeline will also be automated and is discussed in detail in §4.2. We will be developing additional extensions for the intelligent queue so that it can poll the Maunakea weather station, seeing monitor and laser traffic control system [14] as part of its decision making process. We will be coordinating closely with the Mauna Kea Laser Operators Group and the other observatories on the safe use of the UV laser.

## 3. ADAPTIVE OPTICS: IMAGING PERFORMANCE AND ENSQUARED ENERGY

We have maintained detailed error budgets for the expected adaptive optics performance of the prototype Robo-AO system under different observing conditions and have validated the performance on sky [8]. This error budget was originally developed by Richard Dekany (Caltech) and collaborators, and has additionally been validated against the performance of the Keck and 5-m Hale laser AO systems. Using this same tool, we have estimated the performance of the RTS AO system (Table 2). The error budgets use measured Maunakea $C_n^2(h)$ profiles derived from a combination of the Gemini ground-layer study [15] and an analysis, by our team, of the first three years of data from the Maunakea summit MASS/DIMM seeing monitor running since fall 2009. A conservative 0.44″ of additional dome seeing has been added to the $C_n^2(h)$ profiles.

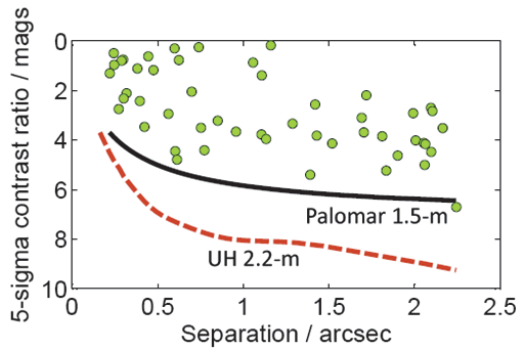

**Figure 3.** Measured visible 5-σ contrast ratio vs. separation for the prototype Robo-AO system in the best 25% conditions (black) and the predicted contrast achievable with the RTS AO system in median, 50%, conditions (dashed red). The green circles represent actual companions detected as part of our survey of 715 Kepler candidate exoplanet host stars [16].

The superior Maunakea atmospheric statistics and higher spatio-temporal order adaptive correction combine to reduce the effective wavefront error in median conditions by nearly 100 nm in quadrature over the prototype at Palomar. This leads to dramatically increased imaging performance in all conditions at red and longer wavelengths: in median conditions, and pointing nearly at zenith, the AO system will deliver an i'-band Strehl ratio of 20% and an image width of 0.08″ vs. 10% and 0.15″ for the prototype. Figure 3 shows the expected 5-σ contrast ratio vs. separation improvement with the RTS AO system. RTS will enable the detection of closer companions and achieve deeper contrasts, e.g. an extra ~1.8 magnitudes at 0.5 arc seconds. The RTS AO will also provide significant correction in median or better conditions for the shortest visible wavelengths. For guide sources brighter than $m_V=17$ (MV), near diffraction-limited-image-width performance is limited by high-order wavefront errors from the laser system.

In addition to traditional measures of adaptive optics performance, we calculated the expected ensquared energy within 0.3″×0.3″

corresponding to 2×2 spaxels of the IFS following the model outlined in [17] that shows accuracy of a few percent for Strehl ratios as low as 4%. Energy in each of the non-diffraction-limited residual halos (scattering, seeing, partially corrected) is proportional to the phase variance of the residual errors and is combined with an appropriately scaled diffraction-limited PSF and blurred by residual tip-tilt errors. Figure 4 shows the expected ensquared energy percentage over the spectral range of the spectrograph in median conditions using a target SNIa as the tip-tilt guide star (with an approximate A0 type spectrum). Tip-tilt sensing is performed in a combined r'+i' filter by the EMCCD camera (the system takes advantage of significant image sharpening at visible wavelengths). The calculated ensquared energy percentages stay relatively constant until residual tip-tilt errors start to dominate the error budget past $m_V$~18.

| Percentile Seeing | 25% | 50% | 50% | 75% |
|---|---|---|---|---|
| Atmospheric $r_0$ | 22.1 cm | 16.8 cm | 16.8 cm | 10.3 cm |
| Effective seeing at zenith (inc. dome seeing) | 0.69" | 0.80" | 0.80" | 1.00" |
| Zenith angle | 15 degrees | 15 degrees | 45 degrees | 35 degrees |
| **High-order Errors** | **Wavefront Error (nm)** | | | |
| Atmospheric Fitting Error | 35 | 39 | 46 | 51 |
| Bandwidth Error | 32 | 37 | 44 | 48 |
| High-order Measurement Error | 32 | 35 | 36 | 41 |
| LGS Focal Anisoplanatism Error | 74 | 102 | 123 | 158 |
| Multispectral Error | 5 | 5 | 94 | 35 |
| Scintillation Error | 12 | 15 | 27 | 27 |
| WFS Scintillation Error | 10 | 10 | 10 | 10 |
| Uncorrectable Tel / AO / Instr Aberrations | 38 | 38 | 38 | 38 |
| Zero-Point Calibration Errors | 34 | 34 | 34 | 34 |
| Pupil Registration Errors | 21 | 21 | 21 | 21 |
| High-Order Aliasing Error | 12 | 13 | 15 | 17 |
| DM Stroke / Digitization Errors | 1 | 1 | 1 | 1 |
| **Total High Order Wavefront Error** | **110 nm** | **134 nm** | **182 nm** | **192 nm** |
| **Tip-Tilt Errors** | **Angular Error (mas)** | | | |
| Tilt Measurement Error | 11 | 12 | 15 | 13 |
| Tilt Bandwidth Error | 8 | 9 | 9 | 10 |
| Science Instrument Mechanical Drift | 6 | 6 | 6 | 6 |
| Residual Telescope Pointing Jitter | 2 | 2 | 2 | 2 |
| Residual Centroid Anisoplanatism | 1 | 1 | 2 | 2 |
| Residual Atmospheric Dispersion | 1 | 1 | 3 | 3 |
| **Total Tip/Tilt Error (one-axis)** | **15 mas** | **16 mas** | **19 mas** | **18 mas** |
| **Total Effective Wavefront Error (IRTT)** | **127 nm** | **150 nm** | **191 nm** | **202 nm** |
| **Total Effective WFE (VISTT)** | **134 nm** | **157 nm** | **195 nm** | **207 nm** |

| Spectral Band | λ | λ/D | Strehl | FWHM | Strehl | FWHM | Strehl | FWHM | Strehl | FWHM |
|---|---|---|---|---|---|---|---|---|---|---|
| g' | 0.47 μ | 0.044" | 7% | 0.06" | 2% | 0.07" | 1% | 0.11" | 0% | 0.48" |
| r' | 0.62 μ | 0.058" | 19% | 0.07" | 10% | 0.07" | 6% | 0.08" | 1% | 0.12" |
| i' | 0.75 μ | 0.070" | 30% | 0.08" | 20% | 0.08" | 14% | 0.08" | 5% | 0.10" |
| J | 1.25 μ | 0.117" | 64% | 0.12" | 54% | 0.12" | 45% | 0.13" | 33% | 0.13" |
| H | 1.64 μ | 0.153" | 76% | 0.16" | 69% | 0.16" | 62% | 0.16" | 51% | 0.16" |

**Table 2.** Example error budget for RTS AO under different seeing conditions ($r_0$) and for different zenith angles (z) assuming an on-axis $m_V$=17 MV star for tip-tilt sensing and on-axis science target. Performance in the visible bands (light blue) uses an infrared tip-tilt signal with a Y+J+H filter. Performance in the infrared bands (red) uses a visible tip-tilt signal with an r'+i' filter. Measurement error arises from finite laser photoreturn, WFS read noise, sky noise, dark current, and other factors. Focal anisoplanatism is an error arising from the finite altitude of the Rayleigh laser resulting in imperfect atmospheric sampling. Multispectral error arises from differential refraction of UV and visible/NIR rays. 'mas' indicates milli arc seconds, (") indicates arc seconds.

## 4. RTS INFRARED INTEGRAL FIELD SPECTROGRAPH (IFS)

We have designed a spectrograph for the RTS that is optimized for quick target acquisition, high throughput, and streamlined calibration and data reduction. The spectrograph makes use of an Integral Field Unit (IFU) that divides the traditional image plane into spatial pixels, "spaxels", where each spaxel is dispersed to generate an individual spectrum on the detector. The RTS integral field spectrograph (IFS) is designed to have an average spectral resolution of R~100 that simultaneously covers the near-infrared band pass from 840 to 1830 nm with a spatial sampling of 0.15″ per spaxel. The primary components of an IFS are: relay optics; a method that samples the image plane (e.g., lenslet array, slicer, fibers); dispersing element; collimator and camera optics; and detector.

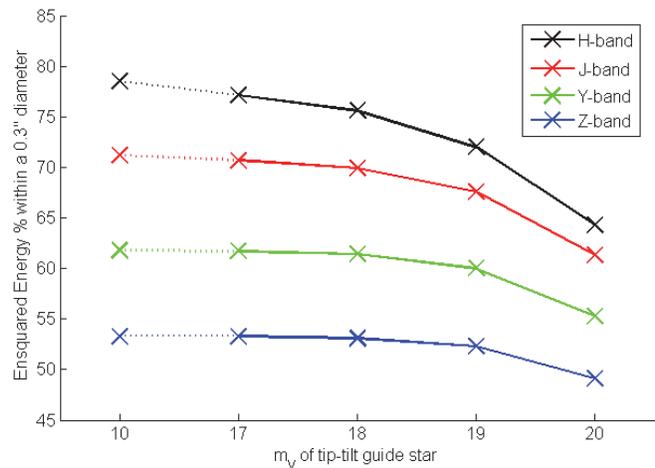

**Figure 4.** Ensquared energy % at different observing bands for a 0.3″ aperture at a Zenith angle of 30°. We use a SNIa target as the tip-tilt guide star. (mv<18.9 out to z~0.1)

Our team has developed a design for the RTS IFS that maximizes the use of the pixels of a Hawaii-2RG (18μm pixel size, 2048x2048) detector, while also allowing for the maximum field of view with a very conservative separation between neighboring spectra to allow for high-fidelity data reduction. The Hawaii-2RG will operate in a separate cryostat from GL Scientific, with a single band pass filter mounted in front of the detector, and use a PB-1 array controller and associated software (developed and built at UH, [18]). The IFS is designed to have no moving opto-mechanical mechanisms (e.g., filter or grating stages) to reduce the overall complexity and cost of the instrument, and to allow for stable spectra.

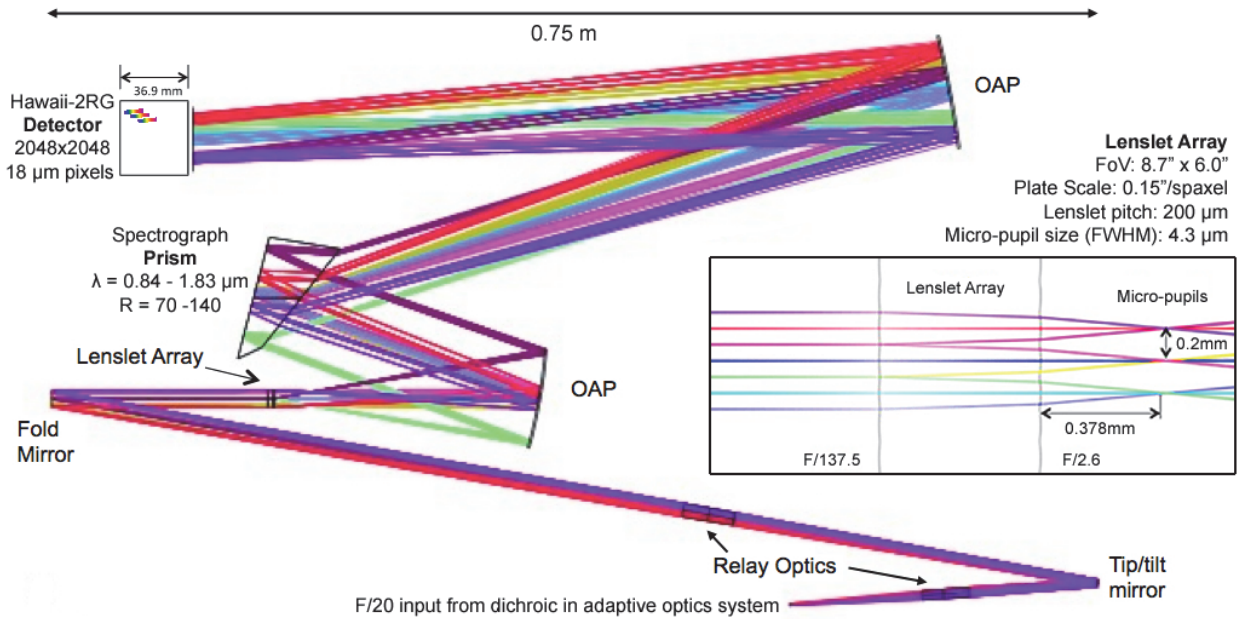

**Figure 5.** Optical diagram of the RTS integral field spectrograph illustrating the primary optical components. The F/20 input from the AO system enters at the bottom and the tip-tilt mirror is placed at the intermediate pupil of the F/137.5 relay optics that generates the 0.15″ plate scale at the location of the 58×40 lenslet array. (**Inset**): The ray trace of 3 adjacent lenslets. The F/137.5 beam enters from the left and enters the left surface of the lenslet that has a thickness of 1mm and the right surface of the lenslet has the majority of the power on the backside to concentrate the light down to well separated pupil images past the array to f/2.6 outgoing beam. Each lenslet produces a micro-pupil that feeds a standard spectrograph that has a single fixed prism (R~70-140) that disperses the near-infrared bandpass (840 - 1830 nm) to 2,320 individual spectra on to a Hawaii-2RG detector.

An optical design layout of the IFS is shown in Figure 5. The IFS accepts the adaptive optics-corrected F/20 beam from the RTS AO system (at the focus of OAP#2, Figure 2, §2.3) and uses an all-refractive, afocal and telecentric relay to reimage the focal plane at F/137.5 on a 58 × 40 lenslet array. The lenset array is designed with the majority of the power on backside of the lens surface to reduce scattered light between adjacent lenslets. The fill factor of the lenslet array is ~95% with a throughput of ~95% using a low-OH fused quartz substrate. Each lenslet then produces a micropupil or "spot" at the focus of a standard Czery-Turner spectrograph that uses a dispersive prism. The optics between the lenslet array and detector are a standard collimator optics, prism, and camera optics. The collimator and camera consists of two concave mirrors with a square f/2.7 beam and f/8.2 beam, respectively. Each spot is dispersed into a spectrum that is 180 pixels in length with a separation between each neighboring spectrum of 5 pixels. In total there are 2,320 spectra on the detector that yield a field of view of 8.7″×6.0″. For stability in the spectra and simplicity of the instrument, the IFU will make use of a single-fixed direct vision prism that disperses across the entire λ = 840 - 1830 nm band pass yielding a range of spectral resolutions of R = 70 – 140. Note that the IFS bypasses the RTS atmospheric dispersion corrector to maintain high throughput and as a result atmospheric dispersion will be corrected in software (§4.2). A direct vision prism will make use of a combination of two glasses (i.e., BaF2/S-FTM16), where the first glass disperses the light and the second piece of glass reflects the light.

All mirror mounts will be made of solid aluminum blocks and the lens mounts will use a spring-loaded ring to axially hold lenses. This allows for differential temperature changes. The alignment procedure for the IFS will make use of tip/tilt of the fold mirrors and adjustments in the radial direction for the lens. The RTS IFS has the distinct advantage of no moving mechanism and directly builds on the heritage and experience of OSIRIS and GPI.

### 4.1 RTS IFS sensitivity calculations

We have developed an exposure time calculator (ETC) for the RTS system. The ETC predicts the sensitivity for a Vega-like point source of a given magnitude, m, at a specified exposure time, t, in standard Z-, Y-, J-, and H-band filters (spanning the full bandpass of the IFS). First, a Vega-like synthetic spectrum (appropriately blue for SNIa) is modified to account for telluric transmission and propagated through the telescope and instrument, accounting for pupil areas and throughputs. For the UH 2.2-m telescope, we assume a telescope throughput of 0.852, and an AO and IFS throughput of 0.75 and 0.35 respectively. We adopt a detector read-noise of 3e- and a dark current of 0.01e-/s. We simulate both seeing-limited and AO-corrected cases, the latter includes AO throughput losses. The source is distributed over lenslet and detector pixels using either a seeing-limited or AO PSF. We assume median seeing conditions at Maunakea (scaled by airmass$^{3/5}$) and include 0.44″ of dome-seeing for an effective seeing at Zenith of FWHM=0.8″. The seeing-limited square-aperture diameter is 0.8″. In the AO corrected case, we use the ensquared energy calculated in §3 and an aperture diameter of 0.3″.

**Table 3.** Seeing vs. AO Signal-to-Noise

| Mag | Y-band SNR AO | Y-band SNR Seeing | J-band SNR AO | J-band SNR Seeing | H-band SNR AO | H-band SNR Seeing |
|---|---|---|---|---|---|---|
| 10 | 12230.7 | 12348.8 | 14186.7 | 13593.9 | 12905.9 | 11891.3 |
| 11 | 7713.7 | 7774.8 | 8941.9 | 8535.6 | 8100.2 | 7233.5 |
| 12 | 4863.1 | 4879.6 | 5631.9 | 5322.7 | 5054.4 | 4230.9 |
| 13 | 3063.1 | 3039.4 | 3540.5 | 3268.5 | 3111.5 | 2331.3 |
| 14 | 1924.9 | 1861.1 | 2215.6 | 1947.5 | 1862.1 | 1199.5 |
| 15 | 1202.9 | 1101.0 | 1371.8 | 1105.6 | 1060.8 | 580.1 |
| 16 | 742.2 | 613.9 | 830.5 | 589.9 | 565.4 | 266.1 |
| 17 | 445.6 | 315.4 | 482.9 | 293.9 | 281.6 | 116.1 |
| 18 | 253.5 | 148.4 | 262.9 | 136.7 | 130.4 | 48.6 |
| 19 | 131.9 | 64.9 | 130.9 | 59.7 | 55.8 | 19.8 |
| 20 | 59.4 | 27.0 | 57.0 | 25.0 | 21.3 | 8.0 |
| 21 | 26.3 | 11.0 | 25.1 | 10.2 | 8.7 | 3.2 |

Note. – SNR is summed over the whole band.

The point source spectrum is integrated over the Z-, Y-, J-, and H-band filters to produce the expected signal in e-/s. A similar integration is done on a background telluric emission spectrum for Maunakea, including lines and continuum, for the same square apertures and filters to produce the expected background signal. Note we assumed a precipitable water vapor of 1.6 mm and an airmass of 1.15 for the background emission spectrum. The signal and background fluxes are multiplied by the integration time. The expected noise is estimated including photon noise from the point source and background flux, and read noise and dark current from the detector. Note, we assume ~60 pixels along the dispersion direction for each band (R=100, Nyquist sampled) and 2 × 2 spaxel apertures for both the seeing-limited and AO-fed cases, equivalent to plate scales of 0.4″ and 0.15″, respectively.

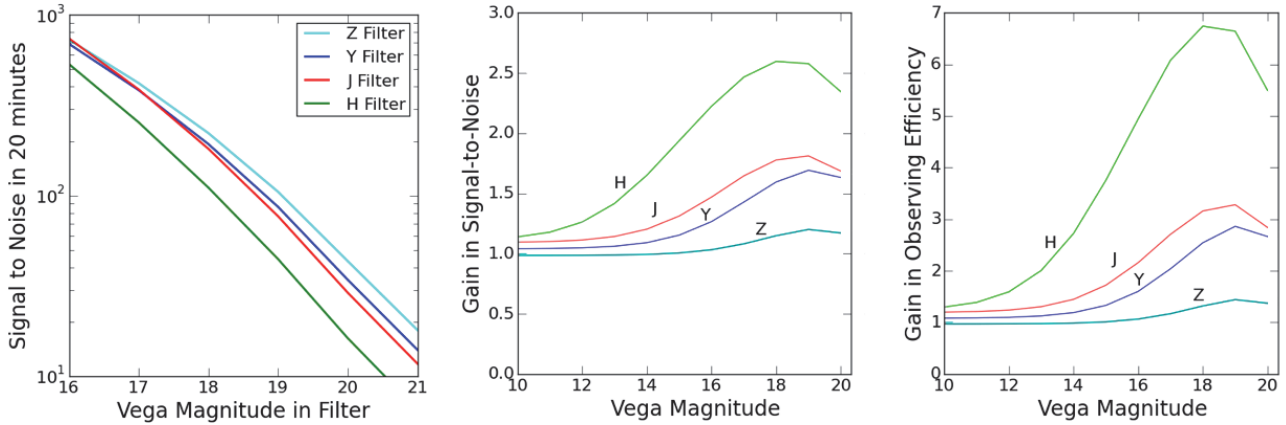

**Figure 6**. **Left:** The expected SNR and gains for the RTS system in the Z-band, Y-band, J-band, and H-band. The SNR is estimated by summing up the flux over the entire R=100 spectrum within each filter. The expected gain in the SNR **(center)** and efficiency **(right)** by including the adaptive optics system in RTS. The gain in efficiency is equivalent to the reduction in integration time needed to reach a specified SNR. We have assumed an integration time of 1200 s, square 0.8″ and 0.3″ apertures in seeing-limited and AO-corrected cases, respectively, and a Zenith angle of 30º.

The expected signal-to-noise (SNR) is greatly enhanced with the addition of an AO system. This is primarily due to the diffraction-limited PSF that the AO system delivers, which allows for smaller extraction apertures and thus smaller backgrounds. A comparison of the expected SNR for both the seeing-limited and AO-fed case is shown in Table 3 and Figure 6 (left). The gain that the AO system provides in both SNR (Figure 6 center) and integration time (Figure 6 right).

To estimate the extinction, $A_{Ks}$, for a SNIa, the filter-integrated photometry can be combined over the Z, Y, J, and H filters. We adopt the following extinction parameterization, which is appropriate for the near-infrared,

$$m_\lambda = m_{\lambda, intrin} + A_\lambda \qquad \text{where} \qquad A_\lambda = A_{Ks}\left(\frac{\lambda}{\lambda_{Ks}}\right)^\alpha$$

and $m_{\lambda,intrin}$ is the intrinsic brightness of the supernova. We adopt $\alpha = -1.95$ according from [19] and assume photometric uncertainties of $\sigma_{m\lambda} = 1/SNR_\lambda$. The resulting random errors on $A_{Ks}$, and corresponding signal-to-noise on the measurement, are shown in Figure 10 for 1200 s exposures for both the seeing limited and RTS adaptive optics corrected cases.

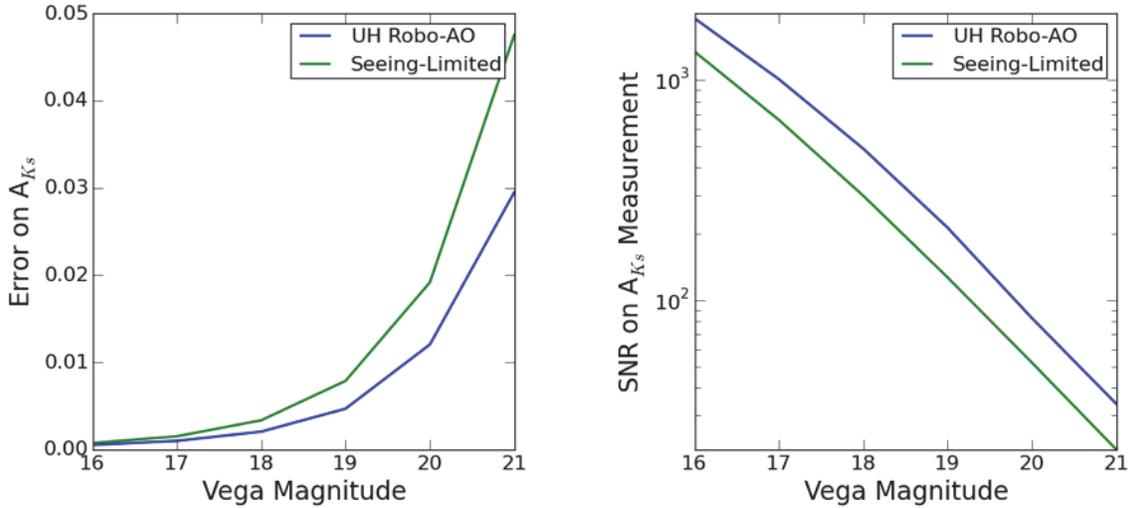

**Figure 7**. Expected random errors **(left)** and SNR **(right)** on the measurement of the extinction, $A_{Ks}$, for the UH RTS (AO-feed) and a seeing-limited instrument. We have assumed a 1200 s integration time, R=100 spectrograph, and wavelength coverage from the Z-band through H-band.

For the range of expected SNIa brightness that will be discovered by ATLAS, the AO correction will increase the λ-weighted SNR on the extinction measurement by approximately 50%, leading to a gain in observing efficiency of greater than 2.

### 4.2 Automated IFS data reduction system

We have purposely designed the RTS spectrograph to have a static instrument configuration and well-separated spectra (5 pixels) to maintain sensitivity and greatly ease the spectral extraction process. We will leverage our experience in the data reduction pipeline development for Keck/OSIRIS to develop an automated data reduction system for the RTS spectrograph. The RTS data reduction system will be based on the Keck/OSIRIS and Gemini/GPI pipelines, and make use of existing software architecture and open source astronomical routines. The RTS-IFS data reduction system will be designed to reduce the spectra in real-time by using a data parser to register written data files (.fits) and associated headers. The data parser software will submit the reduction sequence needed to the data reduction system, as illustrated in Figure 8. We will write custom reduction routines to perform: bad pixels and cosmic ray removal, sky subtraction, wavelength calibration, spectral extraction, atmospheric dispersion correction, telluric correction, and flux calibration.

An internal calibration sources within the RTS AO system (§2.3) will be used to simulate the telescope pupil with an internal thermal source that will be used for flat fielding and spectral extraction, and arc lamps (i.e., Xe, Ar) will be used for wavelength calibration. To generate a data cube, the locations and PSFs of the thousands of spectra are pre-determined from the white light calibration file. This calibration process yields the spectral location and pattern across the detector, and the spectral extraction routine can be performed using a rectangular aperture that traces along the dispersion access (e.g., GPI: [20]). Once each spectrum is extracted, a wavelength solution derived by the arc lamps is implemented aligning each spectrum with a similar wavelength range. Lastly the pipeline will generate a final cube (x, y, λ) that is fully reduced with telluric and flux calibration performed from the calibration database.

We will develop automated tools to detect and extract sources as well as measure the sky background from non-source illuminated spaxels for subsequent subtraction. The 8.7″×6.0″ field of view of the IFS is sufficient for easy acquisition and will encompass the full PSF for stellar extraction routines using aperture photometry per wavelength channel. This stellar extraction routine will generate a calibrated 1d-spectrum for each source detected in field. This will be essential for both subsequent analysis and transfer of data to external users, increasing the impact and immediate usefulness of the data set. We estimate our maximum storage requirements for the spectrograph raw data to be on the order of 15GB per night. Note the prototype AO system already employs a fully automated data reduction pipeline for imaging data [8], saving ~125GB of raw and ~10GB of reduced data per night. All data will be stored and served from UH servers.

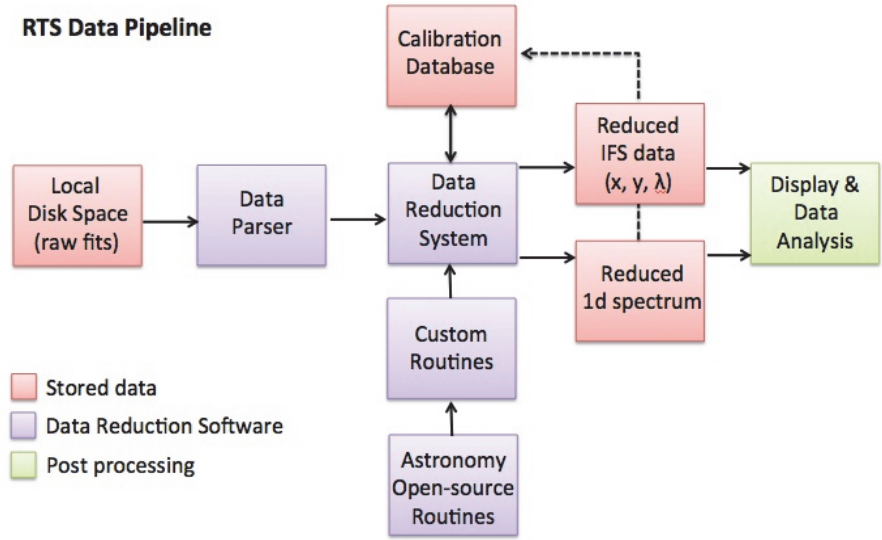

**Figure 8.** Block diagram of the RTS Data Reduction System showing how raw 2D fits files are organized and submitted to the data reduction system and reduced data products stored.

## 5. FROM SNIA DISCOVERY TO OBSERVATIONS AND ANALYSIS

The following flow-chart (Figure 9) shows a conceptual design for the process we will use to discover, classify, follow-up, and analyze SNIa in the survey and derive distances and peculiar velocities. ATLAS automatically creates light curves of non-moving transients, and those that are rising consistent with a SNIa [21] will have observation definition files generated in XML (used by the RTS intelligent queue §2.7) and sent to a SNIa follow-up queue manager with a tolerance of ±1-2 days on the optimal time to observe the SNIa. The queue manager records each transient request and assembles the XML file, light curve, automatic classifications, data, and user comments into a webpage. For targets that have not yet been observed by RTS, the webpage will display the objects in an automated priority order that can be reordered manually. SNIa The webpage will be reviewed daily by at least one member of the science team. Other transients that require rapid observing (e.g. nearby SN, or other high-priority triggers §6.4) can be added manually to the queue manager at any time, or for approved external projects, can be added directly to the 'Community follow-up' program at highest priority.

Once data has been collected for our SNIa follow-up program, the IFS data will be processed with the RTS data pipeline (Figure 8). The reduced spectra will be analyzed for spectral features consisted with a Type Ia, and a crude redshift will be extracted. This data will be combined with catalog redshift data on the host galaxy closest in coordinates to the SNIa and all of the data, including flags for consistency, and estimated redshifts will be updated on the SNIa follow-up queue manager. After it has been determined that we indeed have a SNIa and it is consistent host galaxy redshift, we will extract the slope of the spectrum and correct the peak luminosity measured by ATLAS to determine the un-reddened apparent magnitude which we will then use to determine the host galaxy distance.

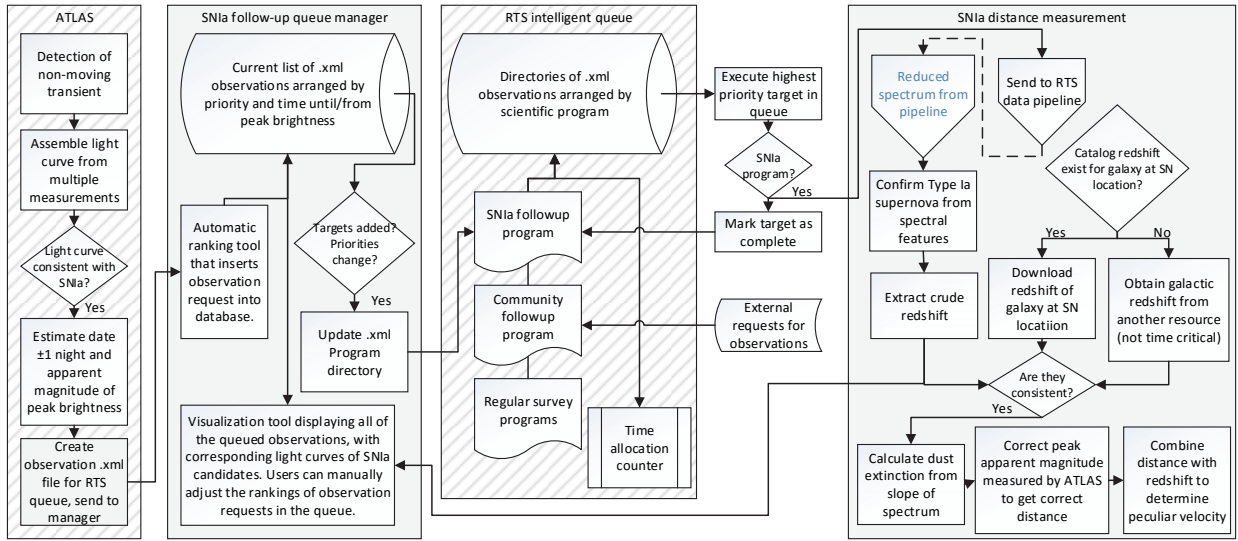

**Figure 9.** Conceptual flow chart for the discovery, classification, follow-up and analysis of SNIa in the survey. Tools to be developed as part of this proposal are shaded in gray, namely the 'SNIa follow-up queue manager', the 'SNIa distance measurement' and the observation definition (XML) generating tool. ATLAS already creates light curves from moving and non-moving transients and will supply the information needed to define an RTS follow-up observation. The RTS intelligent queue is already being used by the prototype AO system, with improvements being made to accommodate community access time. Reduction and delivery of community data are not shown in the figure.

## 6. ADDITIONAL SCIENCE

Beyond the main science proposed above, RTS will enable a broad range of other science cases in the fields of solar system science, exoplanets, star formation, compact objects, and transient phenomena. We highlight a few of these science cases below.

### 6.1 Space Hazards

ATLAS will detect asteroids that pass perilously near the Earth. RTS will enable ATLAS to assess the potential hazard of each asteroid, which is proportional to its total mass. Mass is determined from the combination of reflectivity and density, both of which are very uncertain. For example, fewer than 10% of asteroids are composed of iron, but on impact, small iron asteroids are more dangerous than large low-density stony asteroids. Their albedo and densities can be determined using colors and spectra (see figure 10), but they must be measured quickly and simultaneously because asteroids tumble. This can be done efficiently with RTS. With such data, the ATLAS asteroid discoveries with RTS follow-up can be quickly converted to a mass estimate, which is beneficial for understanding the origins and dynamical families of near-Earth asteroids, as well as assessing the real impact hazard.

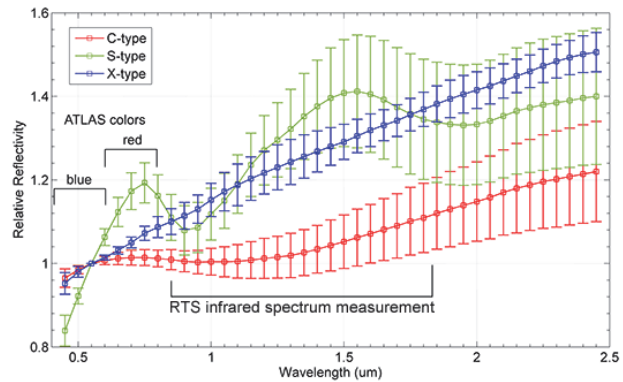

**Figure 10.** Spectra of the most common types of near-Earth asteroids. Colors from ATLAS and a near-infrared spectrum will be able to identify asteroids belonging to the potentially metallic, and therefore most lethal, X-type.

### 6.2 Exoplanets and Their Host Stars

We have entered the era of discovery for large samples of exoplanets. NASA's TESS mission (to launch in 2017; [22]) and the K2 mission extension to Kepler, will detect >5,000 transiting exoplanets. RTS will play a key role in turning these planet candidates into confident detections and to characterize their stellar host systems. An RTS follow-up survey will serve three functions that help maximize the return of TESS and K2: 1) vet for false planet detections, usually from a background eclipsing binary or bound secondary star orbited by a larger planet; 2) search for secondary sources of light in the target aperture that will dilute the transit signal, reducing the measured planet radius; 3) provide statistics on the rates of planet occurrence as a function of host star multiplicity. For example, the Robo-AO prototype searched for blended companions to 3,313 exoplanet host candidates identified by NASA's Kepler mission (Fig. 3, [16, 23, 24]). RTS will have a better contrast and an improved inner working angle and TESS/K2 surveys a closer sample of stars. Thus, RTS will identify secondary light sources at much closer physical distances, e.g., ~2 AU for M dwarfs and ~25 AU for G stars; and is more sensitive to binaries, which are most common at separations of ~50 AU [25].

## 6.3 Imaging Exoplanets and Brown Dwarfs at Wide Separations

RTS will also enable a groundbreaking census of wide-separation young exoplanets. Previous discoveries of wide exoplanets have occurred in a heterogeneous fashion, with a notional frequency of ~1% [26, 27]. The rarity of these objects requires the unparalleled efficiency of RTS in order to survey of thousands of young stars. Such an enormous sample is beyond anything previously done or currently planned with conventional high angular resolution platforms (HST or adaptive optics). From a practical standpoint, the necessary contrasts (brightness ratios of ~100-1000) and relevant angular separations (~1-10") are well matched to RTS. Moreover, access to the southern hemisphere is possible from Maunakea, where young stars are preferentially distributed. The resulting survey will produce a sample of dozens of young gas-giant planets, along with many higher mass brown dwarfs. The spectroscopic capabilities of RTS will enable ready follow-up of new discoveries, as the R~100 spectral resolution is well-matched to the primary spectral diagnostics in young substellar objects (types ranging from late-M to late-T), which are largely shaped by water, methane and condensate (dust) opacity in the Y-, J-, and H-bands. Such a large yield will enable proper statistical measurements of the frequency, mass and separations of this wide population as a function of stellar host mass (e.g. [28, 29]), thereby constraining theories as to their origin.

## 6.4 Open-access Observing Time

Funding permitting, we will make RTS available to the US community through NOAO for both standalone surveys and for rapid follow-up of time-domain projects, using the same trigger system developed for ATLAS, and partnering with such projects as the ANTARES event broker [3] or the 'Global Relay of Observatories Watching Transients Happen' [30] to streamline requests. Other potential triggers include SWIFT-Burst Alert Telescope targets-of-opportunity (ToO) for simultaneous visible/near-infrared follow-up of gamma-ray bursts, and select important discoveries announced to ATEL. Once the predicted optical counterparts to LIGO events (e.g. afterglows of short hard GRBs [31-33], "macronovae" or "kilonovae" which decay on the hours to days timescales [34-37]) are identified with wide-field projects, e.g., with Pan-STARRS [38] or ATLAS, they can be added as the highest priority target in the RTS queue and be potentially observed within minutes. RTS also enables rapid visible and infrared adaptive optics imaging of target-of-opportunity events (e.g. for a transient in a host galaxy – was it a SN or AGN variability?). While this capability exists on the some of the world's largest apertures and HST, these are very limited resources.

Open-access time will be spread evenly over the year using our intelligent queue system, ideal for sustained rapid ToO observing and minimizing the time between discovery and characterization. We will work with NOAO to define the specifics of these observations as parameters like the frequency and duration of community time per night and target prioritization will be program specific. Additionally, RTS will be effective to an elevation of ~30º ensuring access to a majority of the sky from Maunakea, including much of the Southern Hemisphere.

## 7. DISCUSSION

The Robo-AO system architecture has proven itself to be very scientifically productive (e.g., [16, 23, 24, 39-42]). Already, several evolutionary (at Kitt Peak [7]) or derivative systems (e.g., KAPAO at Table Mountain [43, 44], iRobo-AO at the IGO [45], and SRAO at SOAR [46, 47]) are at various levels of commissioning or planning. The Rapid Transient Surveyor will significantly expand on the capabilities of Robo-AO to uniquely deliver rapid high-angular resolution imaging and sensitive NIR spectroscopy. RTS will also serve as a pathfinder for a future implementation on 8-30 m class telescopes - crucial in the era of LSST where larger apertures will be necessary to fully characterize the wealth of new discoveries [48].

## ACKNOWLEDGEMENTS

C.B. acknowledges support from the Alfred P. Sloan Foundation. We thank the following people for developing additional science cases for RTS: Michael Connelley, Donald Hall, Klaus Hodapp, Andrew Howard, Michael Liu, Karen Meech, and Bo Reipurth. We thank Richard Dekany for use of the Wavefront Error Budget Tool.

The Robo-AO system, upon which the RTS AO system is based, was developed by collaborating partner institutions, the California Institute of Technology and the Inter-University Centre for Astronomy and Astrophysics, and with the support of the National Science Foundation under Grant Nos. AST-0906060, AST-0960343, and AST-1207891, the Mt. Cuba Astronomical Foundation and by a gift from Samuel Oschin.